\documentstyle[aps,twocolumn,graphicx,ifthen]{revtex}

\newcommand{\mystyle}{pp}

\newcommand{\myscalebox}[1]{\scalebox{0.4}[0.45]{#1}}
\newcommand{\myscaleboxb}[1]{\scalebox{0.6}[0.6]{#1}}
\newcommand{\myscaleboxc}[1]{\scalebox{0.5}[0.5]{#1}}
\newcommand{\captionA}
{Example of the Cluster-Exact Approximation method. A part of a spin glass
is shown. The circles represent lattice sites/spins. Straight lines represent
ferromagnetic bonds the jagged lines antiferromagnetic interactions. The
top part shows the initial situation. 
The construction starts with the spin at the center. The bottom part 
displays the final stage.
The spins which belong to the cluster carry a plus or minus sign which
indicates how each spin is transformed, so that only ferromagnetic
interactions remain inside the cluster. All other spins cannot be added
to the cluster because it is not possible to multiply them by $\pm 1$
to make all adjacent bonds positive. Please note that many other combinations
of spins can be used to build a cluster without frustration.}

\newcommand{\captionB}
{Average ground-state energy of 2d $\pm J$ Ising spin glass for 
linear dimensions $5\le L \le 40$ and FSS fit of form $e(L)=e_{\infty}+
e_1 L^{-e_2}$ resulting in $e_{\infty}=-1.4015(3)$.}

\newcommand{\captionC}
{Distribution of overlaps $P(|q|)$ for ground
states of 2d $\pm J$ Ising spin glass for $L=10,20,40$. Only for large
values of $q$ a difference is visible, so even for large systems there
is a finite probability of overlap $q=0$. The lines are guides for the 
eyes only.}

\newcommand{\captionD}
{Variances $\sigma^2(|q|)$ 
of the distributions of overlaps $P(|q|)$ as function of $L$
for $5\le L \le 40$. For $L\to\infty$ the value converges to a small 
value. A fit of the form $s_{\infty}+s_1 L^{-s_2}$
gives $s_{\infty}=0.004(8)$. The straight line shows a fit with 
$s_{\infty}\equiv 0$, which should be true for a simple ground-state 
landscape.}

\newcommand{\captionE}
{Fraction $X_{q_0}$ of the distributions of overlaps $P(|q|)$ below $q_0=0.2$ 
as function of $L$ for $5\le L \le 40$. The value of $X_{0.2}$ does not
decrease with growing $L$ indicating a complex ground-state landscape.}

\newcommand{\captionF}
{Distribution $P(\delta q)$ for different system sizes $L=10,20,40$ 
where $\delta q=q_2-q_1$ and $q_1\le q_2\le q_3$ are
triplets of absolute values of overlaps from 
independent triplets of ground states.
Only triplets with $q_3 \in [0.25,0.35]$ are used. For a infinite ultrametric
system $\delta q=0$ holds. The distribution turns out to be
independent of the system size indicating the absence
ultrametricity for the ground states. The inset shows the 
average value of $\delta q$ as function of system size $L$.}

\newcommand{\captionG}
{
Distribution $P_{2-fix}(q)$ for different system sizes $L=10,20,40$ 
where $q\in\{q_1,q_2,q_3\}$ and $q_1\le q_2\le q_3$ are
triplets of overlaps from independent triplets of ground states. 
Only $q$-values of triplets are used where the two other overlaps 
are within the
interval $[0.25,0.35]$. Then for a infinite ultrametric system 
$q>0.25$ should hold, while for a metric system just $q>-0.5$ must hold. 
The small inset shows the value integrated from $-1$ to $0.25$. 
With increasing system
size the fraction of the distribution below $0.25$ remains constant although
the distribution itself changes. The lines are guides for the 
eyes only.
}

\newcommand{\captionH}
{Integrated value $I_L$ of $P_{2-fix}(q)$ outside $[q_{fix},q_{EA}]$ 
as function of system size $L$ 
where $q\in\{q_1,q_2,q_3\}$ and $q_1\le q_2\le q_3$ are
triples of overlaps from independent triples of ground states.
Only $q$-values of triples where the two other overlaps are within the
interval $[0.25,0.35]$ are used. 
With increasing system
size the fraction of the distribution outside $[0.25,0.5]$ remains
mainly constant, which indicates that 2d $\pm J$ spin glasses are
not ultrametric.}

\newcommand{\captionI}
{Position of the peak of the distribution of overlaps $P(|q|)$ as function
of system size for $5\le L\le 40$. A fit function of the form $q_{\max}(L)=
q_{EA}+q_1 L^{-q2}$ is given. The Edwards-Anderson order parameter
is obtained by $L\to\infty$ as $q_{EA}=0.50(9)$.}

\newcommand{\captionK}
{The simplest frustrated system: a triple of spins, each pair of spins
connected by antiferromagnetic bonds (dashed lines). It is not
possible to satisfy all bonds.
}

\newcommand{\captionALGO}
{Genetic Cluster-exact Approximation.
}

\newcommand{\captionALL}
{}

\newcommand{\figA}{
\begin{figure}[htb]
\begin{center}
\myscaleboxc{\includegraphics{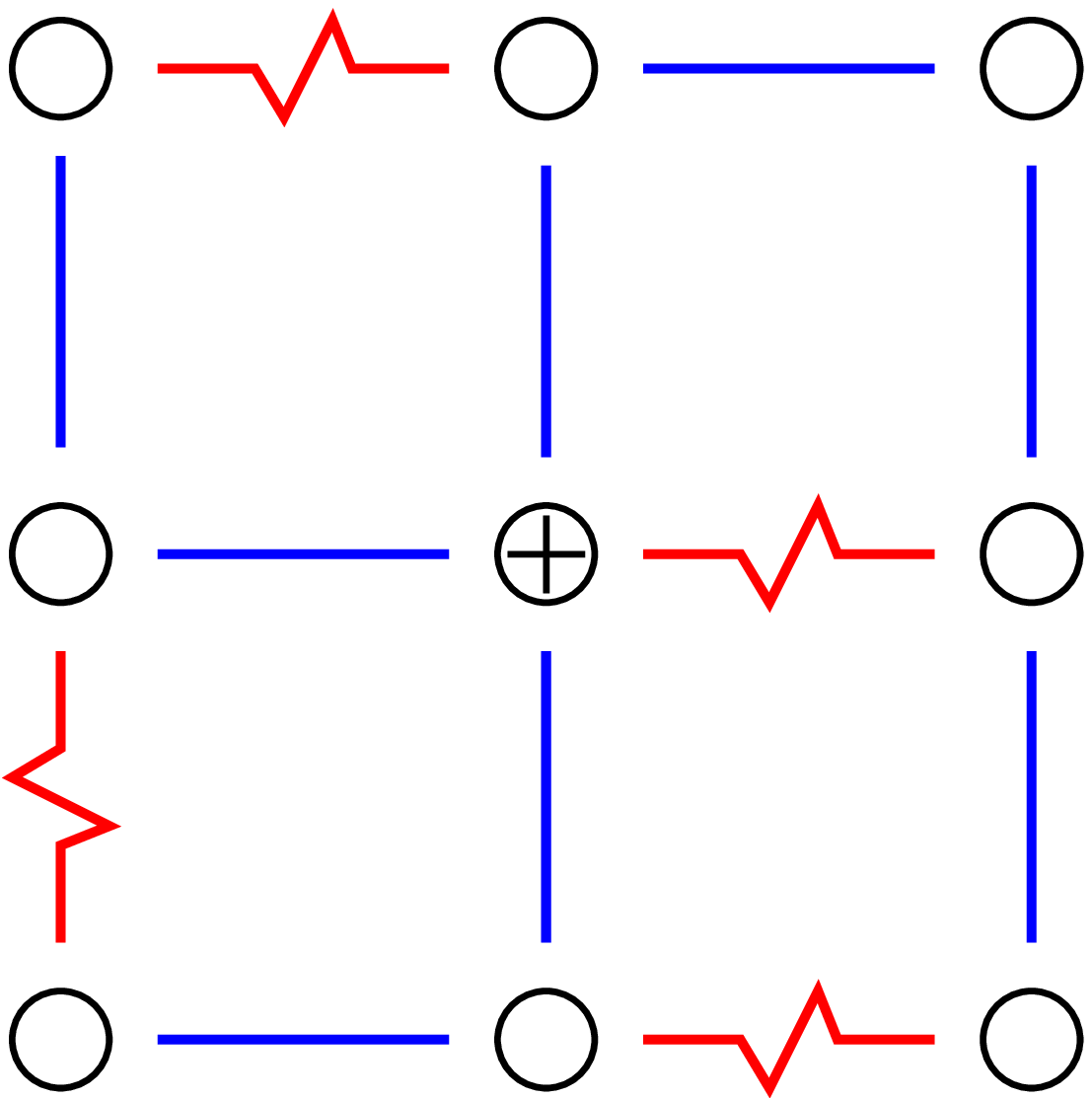}}

\vspace{0.2cm}

\myscaleboxc{\includegraphics{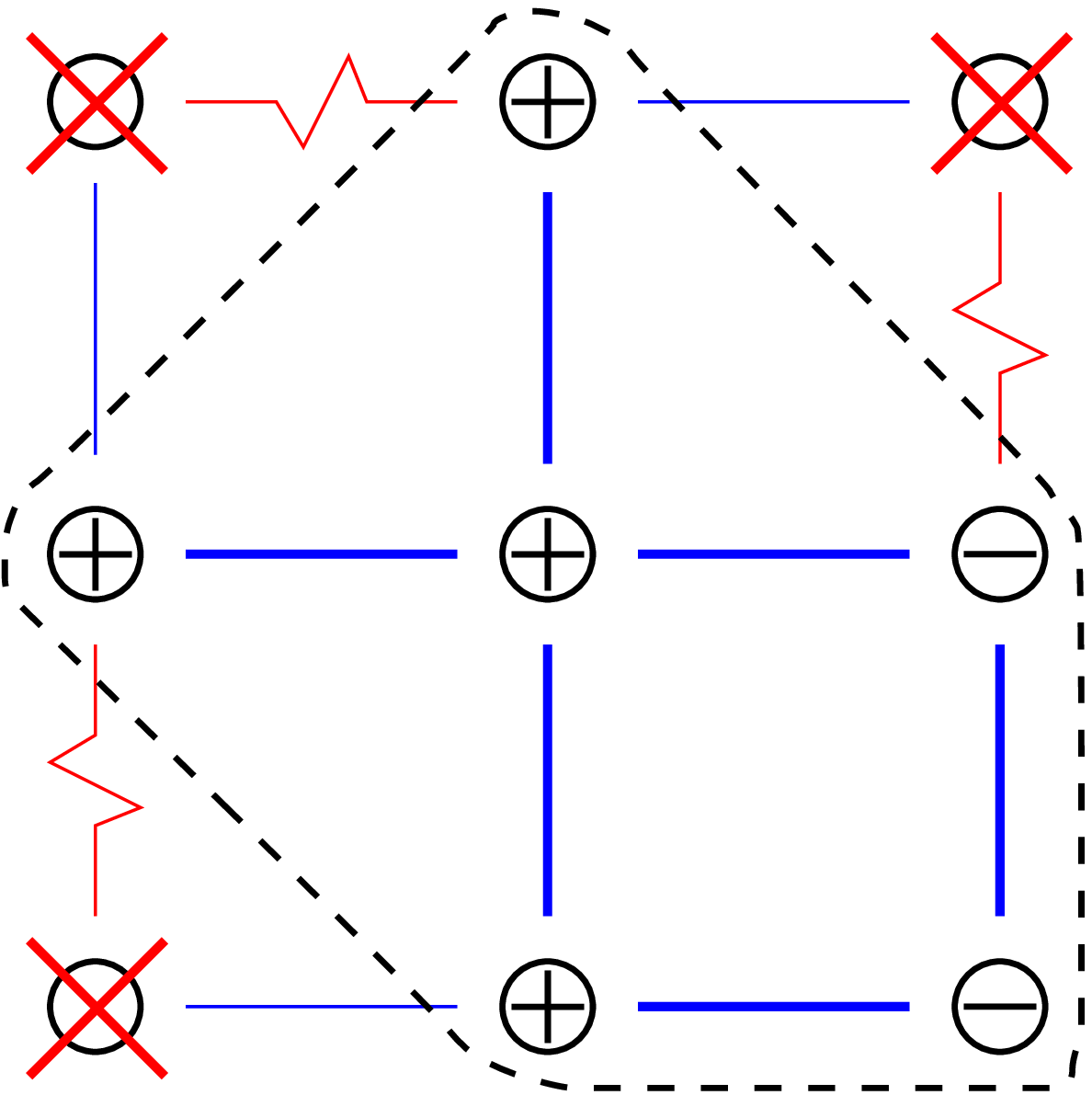}}
\end{center}
\ifthenelse{\equal{\mystyle}{pp}}{\caption{\captionA}}{\caption{\captionALL}}
\label{fig_cea_example}
\end{figure}
}

\newcommand{\figB}{
\begin{figure}[htb]
\begin{center}
\myscalebox{\includegraphics{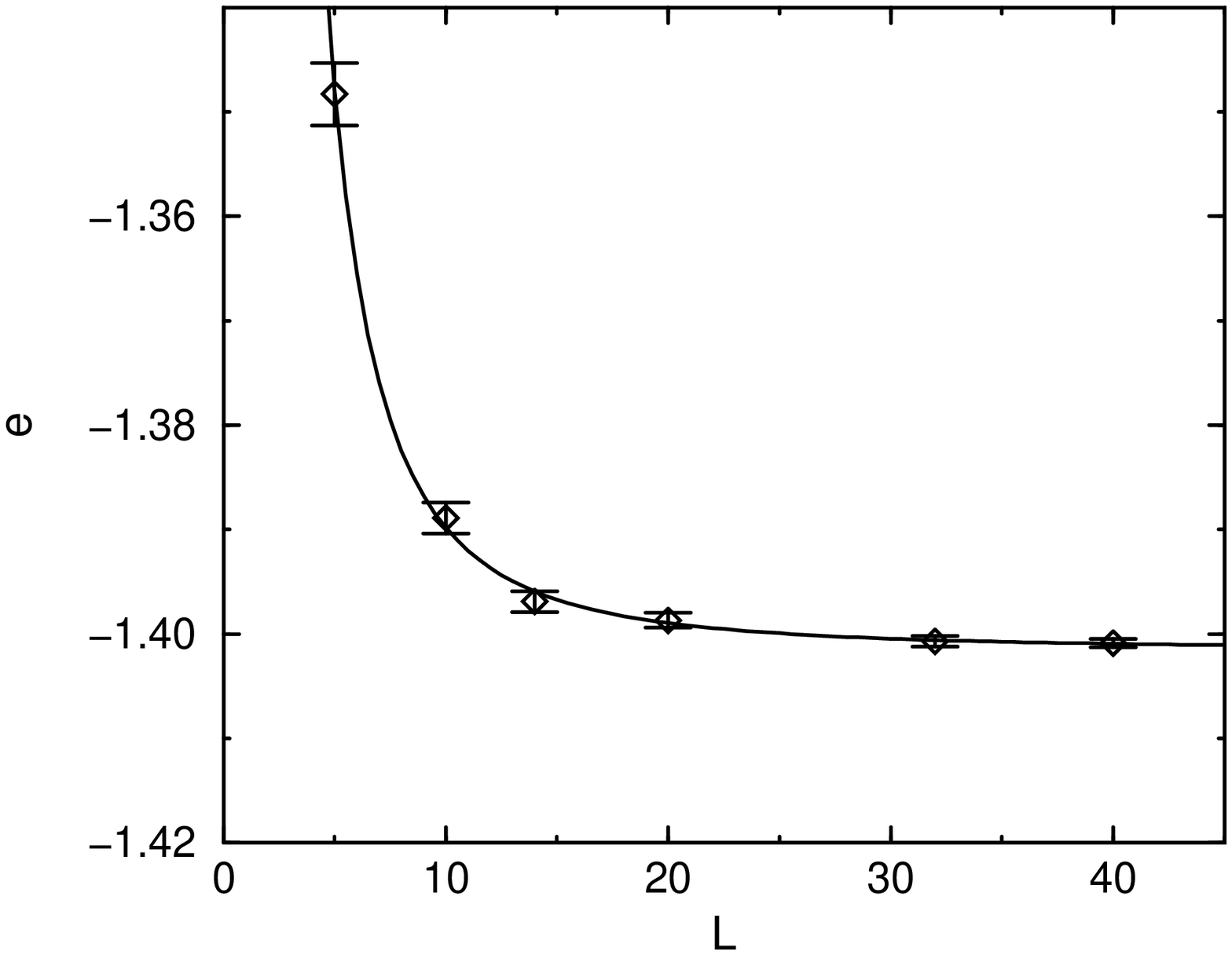}}
\end{center}
\ifthenelse{\equal{\mystyle}{pp}}{\caption{\captionB}}{\caption{\captionALL}}
\label{fig_e_min}
\end{figure}
}

\newcommand{\figC}{
\begin{figure}[htb]
\begin{center}
\myscalebox{\includegraphics{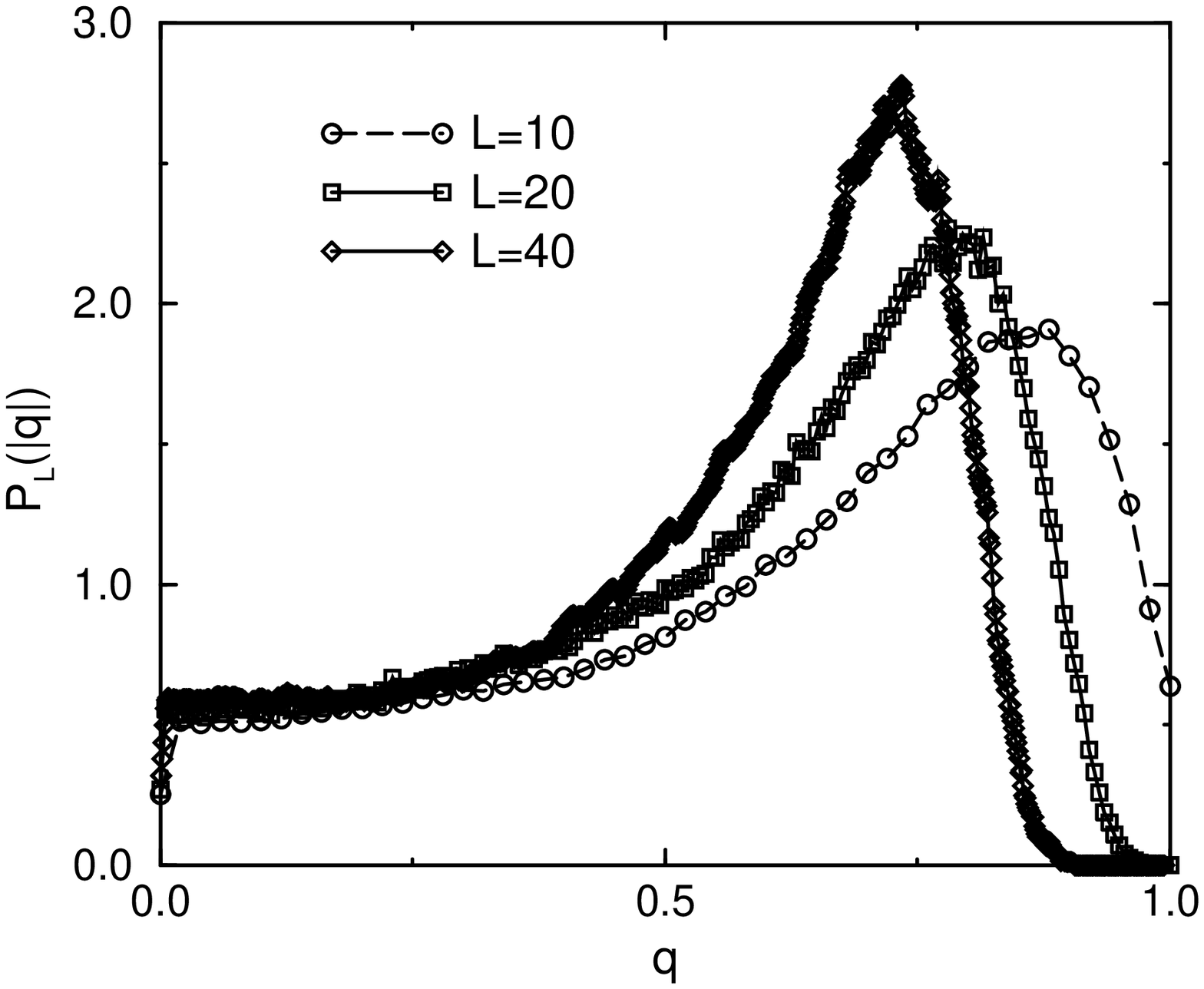}}
\end{center}
\ifthenelse{\equal{\mystyle}{pp}}{\caption{\captionC}}{\caption{\captionALL}}
\label{fig_PLq}
\end{figure}
}

\newcommand{\figD}{
\begin{figure}[htb]
\begin{center}
\myscalebox{\includegraphics{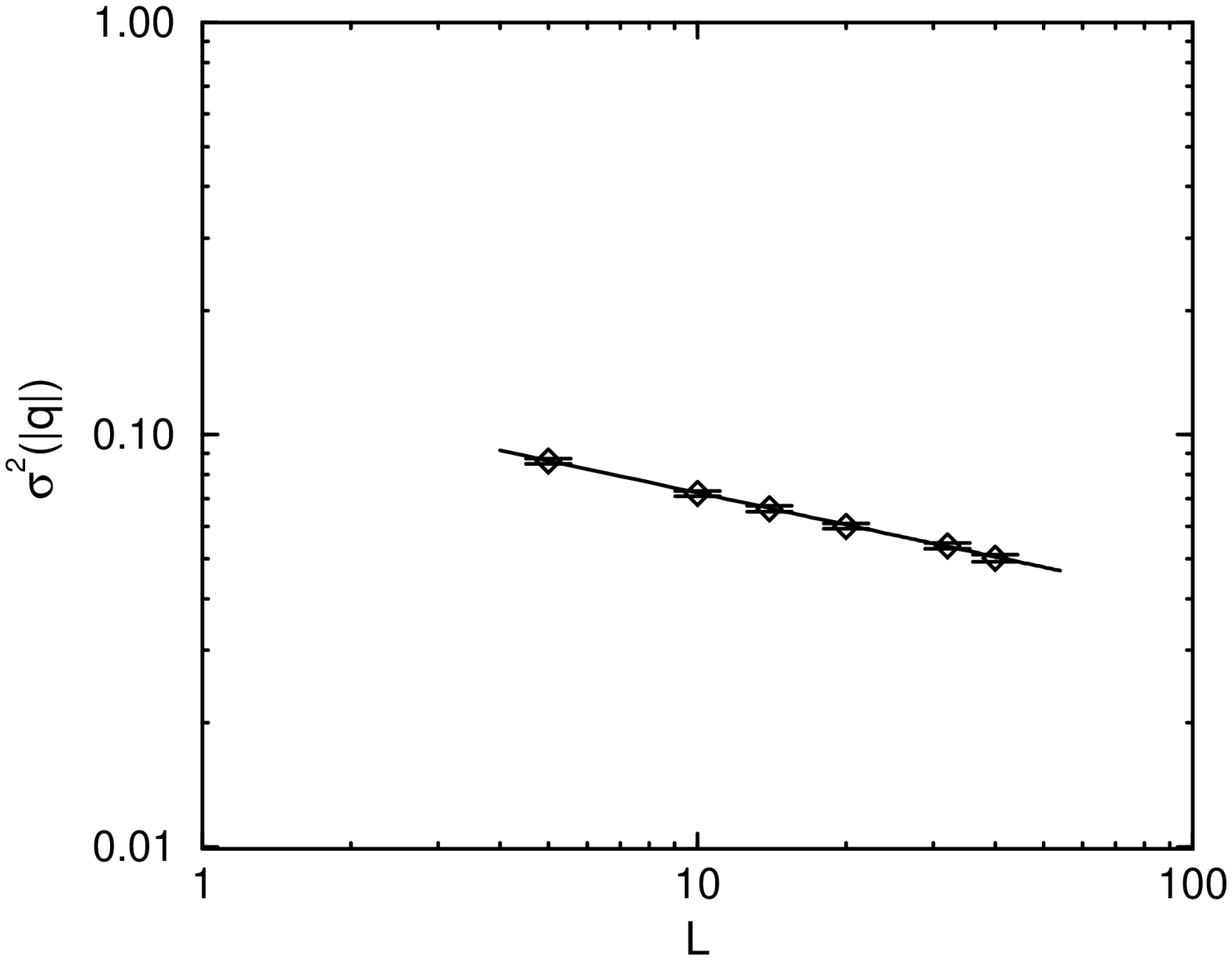}}
\end{center}
\ifthenelse{\equal{\mystyle}{pp}}{\caption{\captionD}}{\caption{\captionALL}}
\label{fig_sigma}
\end{figure}
}

\newcommand{\figE}{
\begin{figure}[htb]
\begin{center}
\myscalebox{\includegraphics{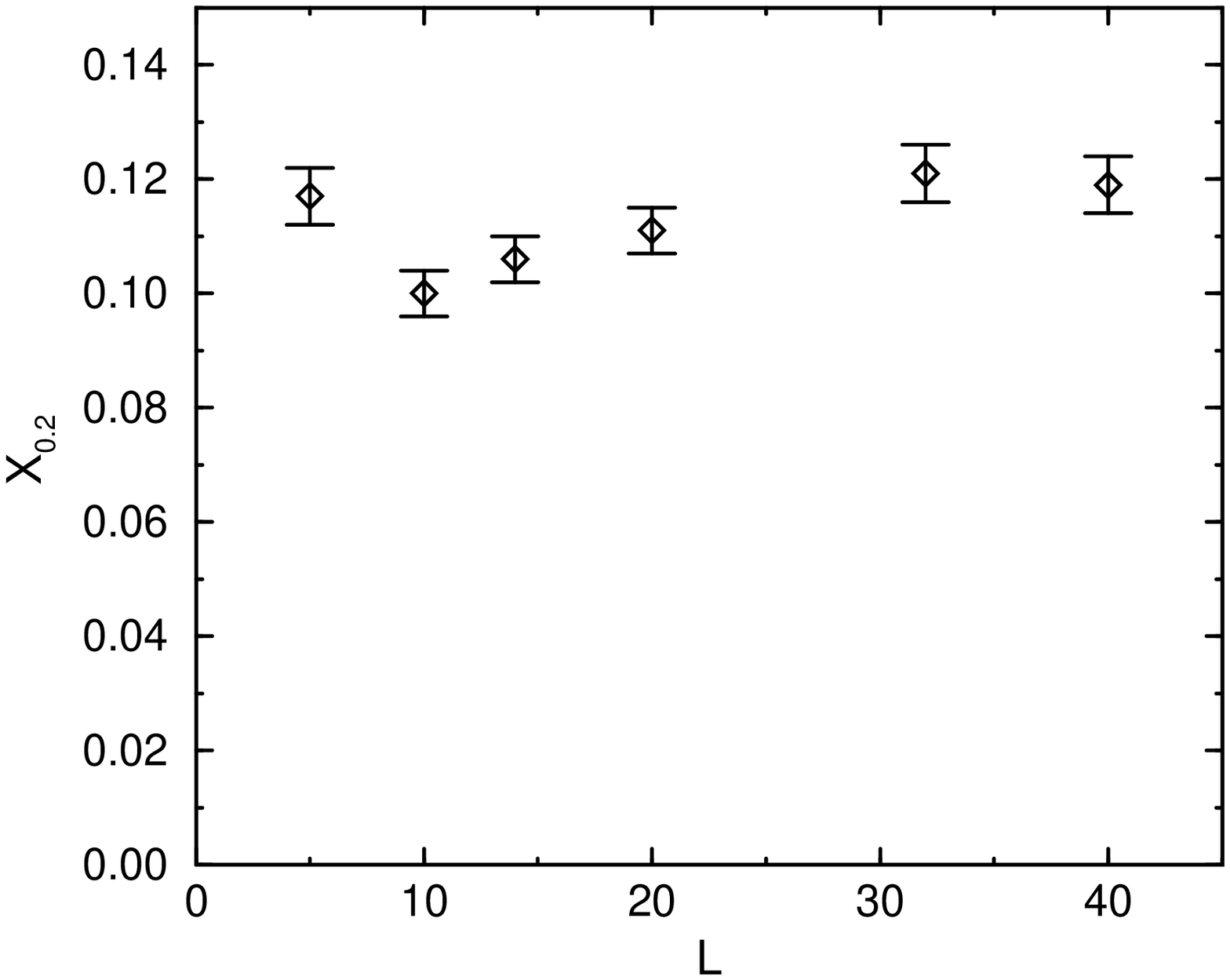}}
\end{center}
\ifthenelse{\equal{\mystyle}{pp}}{\caption{\captionE}}{\caption{\captionALL}}
\label{fig_x02}
\end{figure}
}

\newcommand{\figF}{
\begin{figure}[htb]
\begin{center}
\myscalebox{\includegraphics{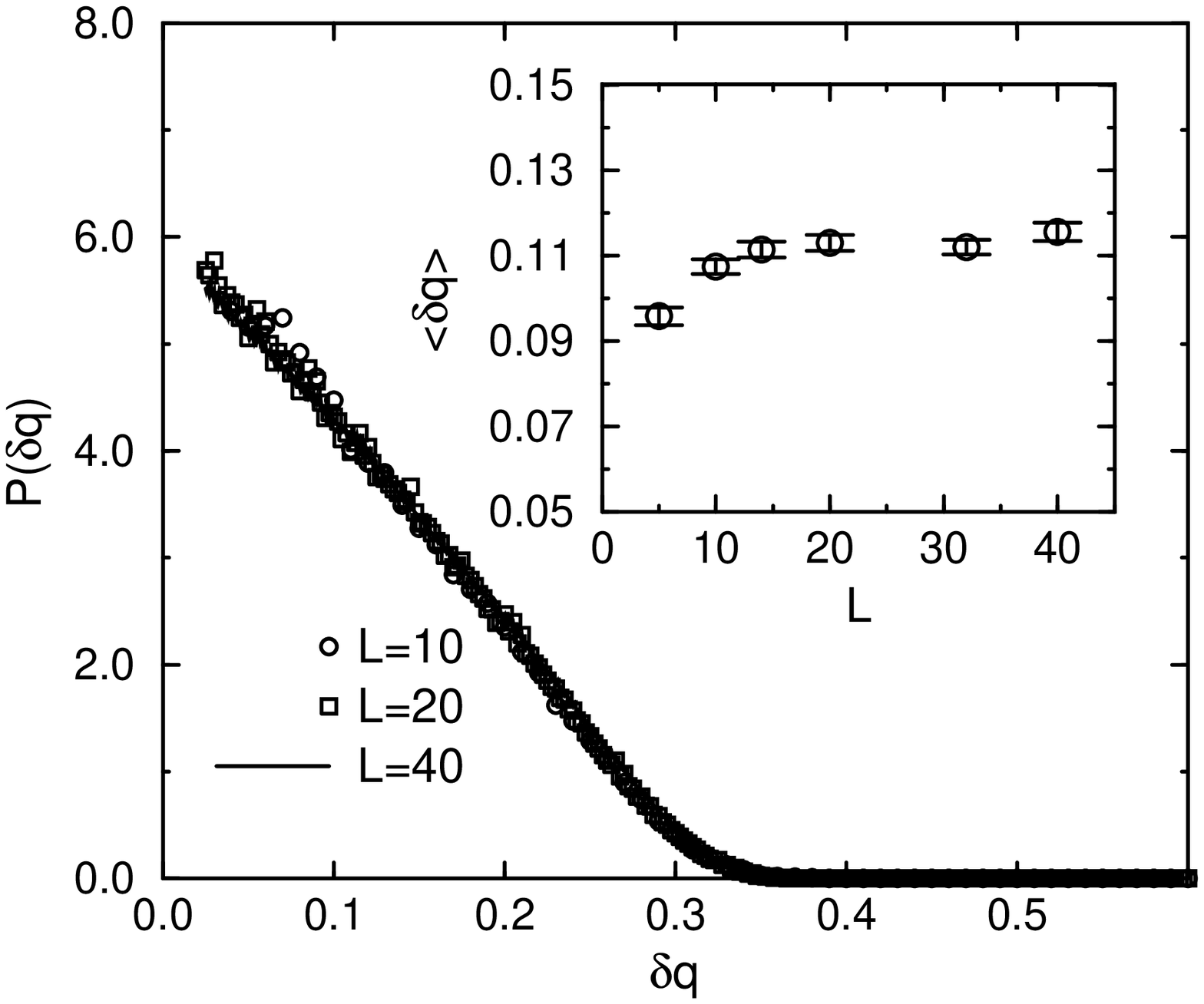}}
\end{center}
\ifthenelse{\equal{\mystyle}{pp}}{\caption{\captionF}}{\caption{\captionALL}}
\label{fig_deltaq}
\end{figure}
}

\newcommand{\figG}{
\begin{figure}[htb]
\begin{center}
\myscalebox{\includegraphics{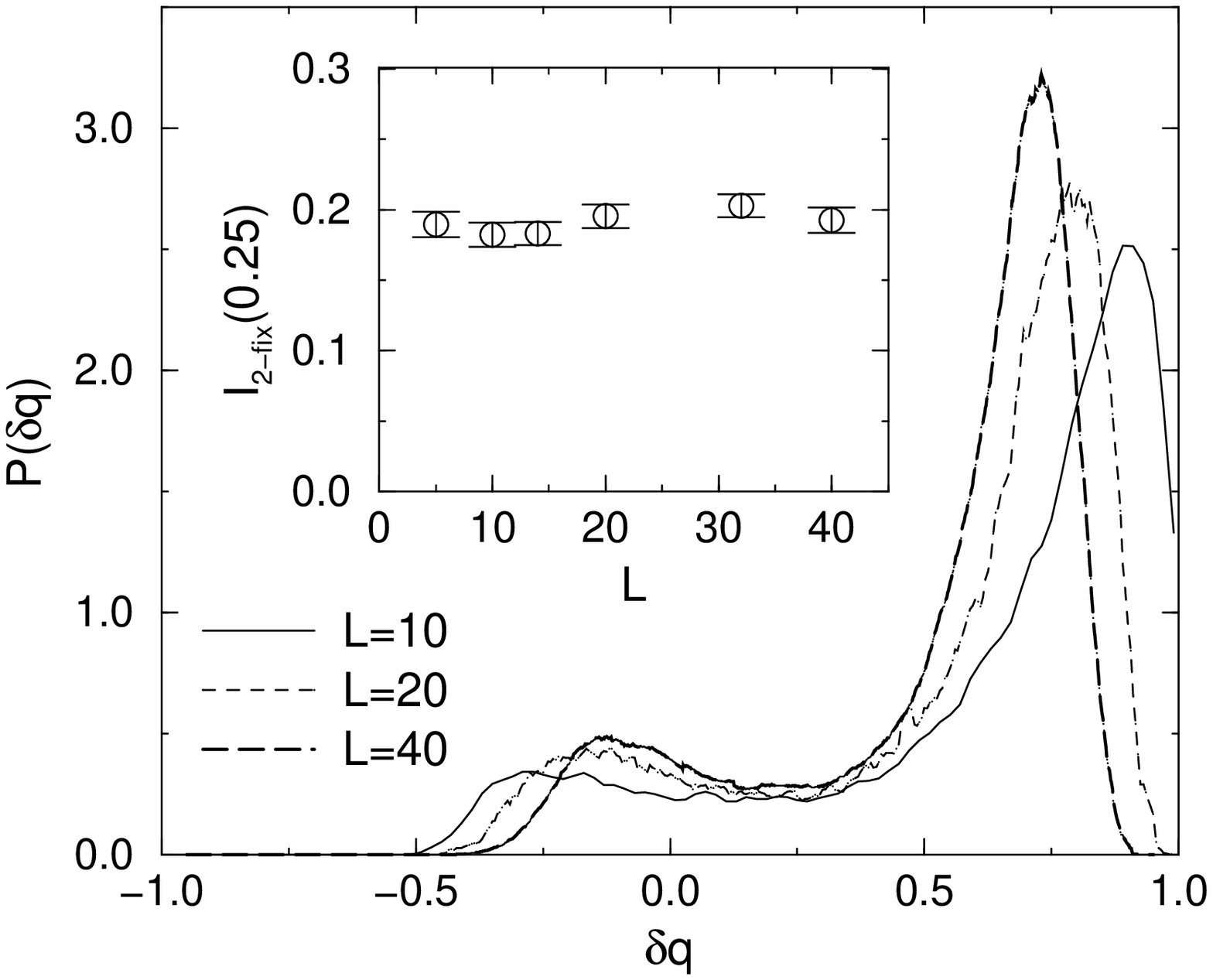}}
\end{center}
\ifthenelse{\equal{\mystyle}{pp}}{\caption{\captionG}}{\caption{\captionALL}}
\label{fig_P_q_2fix}
\end{figure}
}

\newcommand{\figH}{
\begin{figure}[htb]
\begin{center}
\myscalebox{\includegraphics{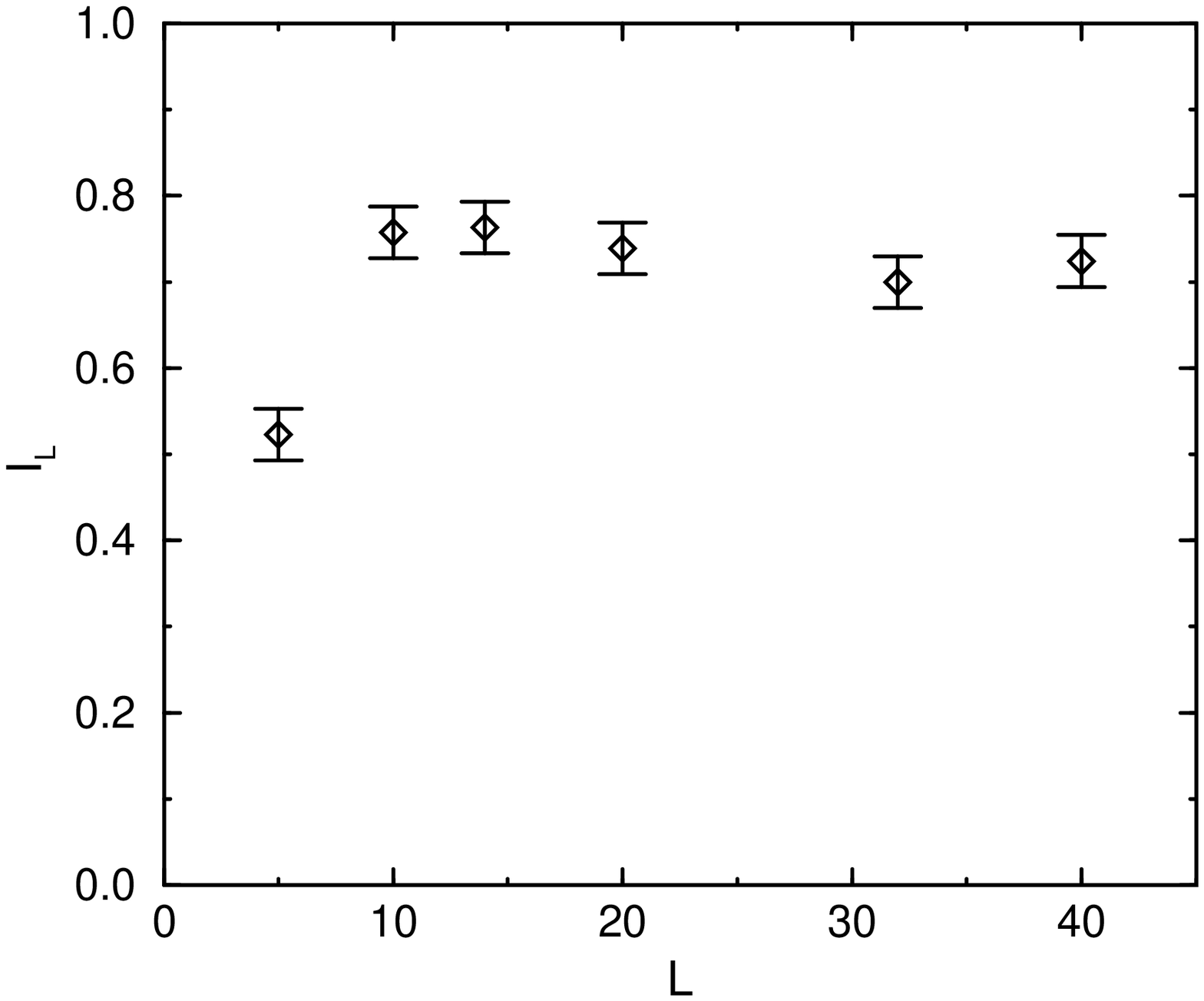}}
\end{center}
\ifthenelse{\equal{\mystyle}{pp}}{\caption{\captionH}}{\caption{\captionALL}}
\label{fig_I_L}
\end{figure}
}

\newcommand{\figI}{
\begin{figure}[htb]
\begin{center}
\myscalebox{\includegraphics{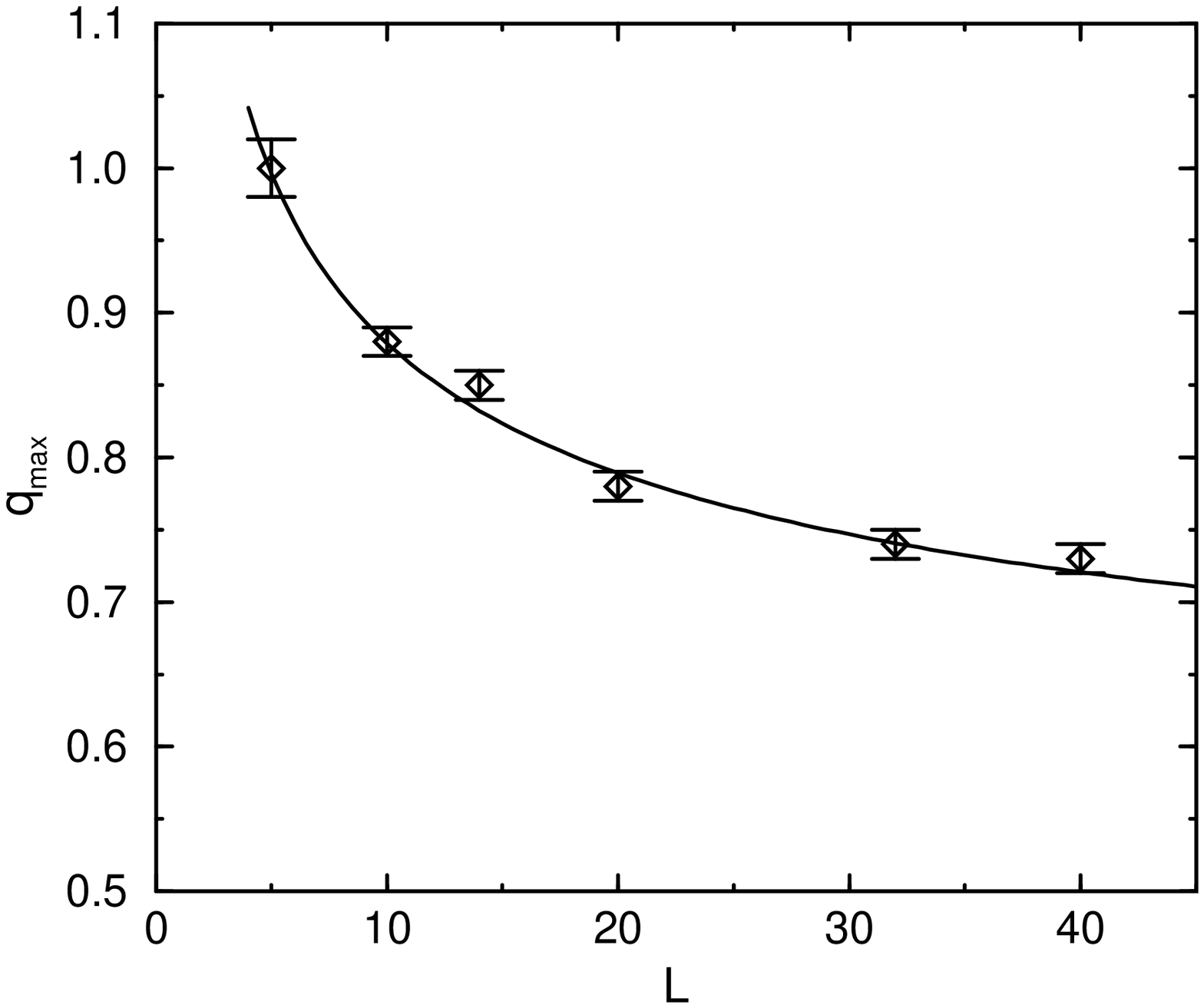}}
\end{center}
\ifthenelse{\equal{\mystyle}{pp}}{\caption{\captionI}}{\caption{\captionALL}}
\label{fig_q_max}
\end{figure}
}

\newcommand{\figK}{
\begin{figure}[htb]
\begin{center}
\myscaleboxb{\includegraphics{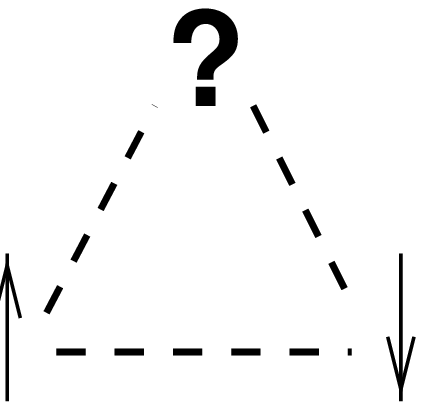}}
\end{center}
\ifthenelse{\equal{\mystyle}{pp}}{\caption{\captionK}}{\caption{\captionALL}}
\label{fig_triangle}
\end{figure}
}

\newcommand{\figALGO}{
\newlength{\mpwidth}
\setlength{\mpwidth}{\textwidth}
\addtolength{\mpwidth}{-2cm}
\begin{figure}

\begin{center}

\begin{minipage}[b]{\mpwidth}
\newlength{\tablen}
\settowidth{\tablen}{xxx}
\newcommand{\tabspace}{\hspace*{\tablen}}
\begin{tabbing}
\tabspace \= \tabspace \= \tabspace \= \tabspace \= \tabspace \=
\tabspace \= \kill
{\bf algorithm} genetic CEA($\{J_{ij}\}$,
$M_i$, $n_o$, $p_m$, $n_{\min}$)\\
{\bf begin}\\
\> create $M_i$ configurations randomly\\
\> {\bf while} ($M_i > 4$) {\bf do}\\
\> {\bf begin}\\
\> \> {\bf for} $i=1$ {\bf to} $n_o \times M_i$ {\bf do}\\
\>\> {\bf begin}\\
\>\>\> select two neighbors \\
\>\>\> create two offsprings using triadic crossover\\
\>\>\> do mutations with rate $p_m$\\
\>\>\> {\bf for} both offsprings {\bf do}\\
\>\>\> {\bf begin}\\
\>\>\>\> {\bf for} $j=1$ {\bf to} $n_{\min}$ {\bf do}\\
\>\>\>\> {\bf begin}\\
\>\>\>\>\> construct unfrustrated cluster of spins\\
\>\>\>\>\> construct equivalent network\\
\>\>\>\>\> calculate maximum flow\\
\>\>\>\>\> construct minimum cut\\
\>\>\>\>\> set new orientations of cluster spins\\
\>\>\>\> {\bf end}\\
\>\>\>\> {\bf if} offspring is not worse than related parent \\
\>\>\>\> {\bf then}\\
\>\>\>\>\> replace parent with offspring\\
\>\>\> {\bf end}\\
\>\> {\bf end}\\
\>\> half population; $M_i=M_i/2$\\
\> {\bf end}\\
\> {\bf return} one configuration with lowest energy\\
{\bf end}
\end{tabbing}

\end{minipage}
\end{center}
\ifthenelse{\equal{\mystyle}{pp}}{\caption{\captionALGO}}
                                 {\caption{\captionALL}}
\label{fig_algo}
\vspace{0.5cm}
\end{figure}
}

\newcommand{\tabA}
{
\begin{center}
\begin{tabular}{ccccc}
\hline
$L$ & $M_i$ & $n_o$ & $n_{\min}$ & $\tau$ (sec) \\ \hline
5 & 8 & 1 & 1 & 0.02 \\
10 & 16 & 1 & 2 & 0.4 \\
14 & 16 & 4 & 2 & 3 \\
20 & 32 & 8 & 2 & 30 \\
32 & 128 & 8 & 2 & 780 \\
40 & 512 & 8 & 2 & 5400  
\end{tabular}
\end{center}

\vspace{0.2cm}

{Tab 1. Simulation parameters: $L$ = system size, $M_i$ = initial size of
population, $n_o$ = average number of offsprings per configuration, $n_{\min}$
= number of CEA minimization steps per offspring, $\tau$ = average computer
time per ground state on a 80MHz PPC601.}

\vspace{0.2cm} 

}

\begin{document}
\title{Ground-state landscape of 2d $\pm J$ Ising spin glasses}

\author{Alexander K. Hartmann\\
{\small  hartmann@theorie.physik.uni-goettingen.de}\\
{\small  Institut f\"ur theoretische Physik, Heidelberg, Germany }\\
{\small and}\\
{\small Institut f\"ur theoretische Physik, Bunsenstr. 9}\\
{\small 37073 G\"ottingen, Germany\thanks{present address}}\\
{\small Tel. +49-551-399568, Fax. +49-551-399631}}

\date{\today}
\maketitle
\begin{abstract}
Large numbers of ground states of  two-dimensional
Ising spin glasses with periodic boundary conditions in both
directions are calculated for sizes up to $40^2$. A combination 
of a genetic algorithm and Cluster-Exact Approximation is used. 
For each quenched realization of the bonds up to 40 independent
ground states are obtained.

For the infinite system a ground-state energy of $e=-1.4015(3)$ is
extrapolated.
The ground-state landscape is investigated using a finite-size scaling
analysis of the distribution of
overlaps. The mean-field picture assuming a complex landscape
describes the situation better than the droplet-scaling model, where for
the infinite system mainly two ground states exist. Strong evidence is
found that the ground states are not organized in an ultrametric fashion
in contrast to previous results for three-dimensional spin glasses.

\end{abstract}

\pacs{75.10.Nr, 75.40.Mg, 02.10.Jf}

\paragraph*{Introduction}

In this work two-dimensional Edwards-Anderson (EA) $\pm J$
spin glasses \cite{binder86} are  investigated. They consist of $N$ spins 
$\sigma_i = \pm 1$, described by the Hamiltonian
\begin{equation}
H \equiv - \sum_{\langle i,j\rangle} J_{ij} \sigma_i \sigma_j
\end{equation}

The spins are placed on a two-dimensional (d=2) 
square lattice of linear size $L$ with periodic boundary conditions in
both directions.
Systems with quenched disorder of the nearest-neighbor interactions (bonds)
are investigated. Their possible values are $J_{ij}=\pm 1$ with equal
probability. A constraint is imposed, so that 
$\sum_{\langle i,j\rangle} J_{ij}=0$.

A question which has led to much controversy is, whether many pure states exist
for finite dimensional, i.e. 
realistic spin glasses. For the infinite ranged Sherrington-Kirkpatrik
(SK)   Ising spin glass \cite{sherrington75}
this question was answered positively by the
replica-symmetry-breaking mean-field (MF) scheme by Parisi
\cite{parisi2}.
Additionally this solution of the
SK-model exhibits \cite{mezard84,franz92}  an
{\em ultrametric } structure: The distances $d^{\alpha\beta}$ between the 
states do not only satisfy the triangular inequality
$d^{\alpha\beta} \le d^{\alpha\gamma} + d^{\gamma\beta}$
but the stronger ultrametric inequality
$d^{\alpha\beta} \le \max(d^{\alpha\gamma}, d^{\gamma\beta})$ as well.
For an introduction to ultrametricity see \cite{rammal86}.
Numerical work on the subject for the SK-model can be found in
\cite{parga84,bhatt86}.

A complete different model is proposed by
the Droplet Scaling (DS) theory 
\cite{mcmillan,bray,fisher1,fisher2,bovier}.
It suggests that only two pure states (related by a global flip) exist and
that the most relevant excitations are obtained by reversing large
domains of spins (the droplets). 
Some authors \cite{newman} disbelieve that the MF theory is consistent
for realistic spin glasses but the question about the existence of many pure
states is not answered. 

In this work the point is not addressed by investigating spin glasses
 at finite temperature, but a different approach is used: an analysis
of ground states is performed. 
The existence of many pure states implies that
between two ground-state configurations 
arbitrary differences are possible. Otherwise
two ground states would only differ by the spin orientations
in some finite domains,
which is always possible in the $\pm J$ model because of the discrete
structure of the interaction distribution. Such a simple structure was
found for example for the 3d random-field Ising model \cite{alex-daff2}.
An investigation whether the ordering of the ground states is
ultrametric is of interest on its own: it is conjectured that an
  ultrametric structure prohibits that the usual
cluster algorithms for simulating a system at {\em finite temperature} work 
efficiently \cite{persky96}.

Next a short overview about numerical results of
finite-dimensional systems for investigating the low-temperature
behavior are given.
In four dimensions the MF picture is well established. Even  
numerical evidence for ultrametricity 
at finite temperature but below the transition temperature $T_G$ was
found \cite{cacciuto97,marinari98}.

For three-dimensional systems recently evidences for the validity of some
basic features of the
MF picture were found using simulations at finite temperature
\cite{marinari96} and ground-state calculations \cite{alex-sg2}.
First attempts
to find ultrametricity by simulation at
finite temperature are given in \cite{sourlas84,caracciolo90}. Evidences
for an ultrametric ground-state landscape of 3d systems were recently
found \cite{alex-ultra}.

The higher-dimensional systems exhibit a spin-glass phase at nonzero
temperature \cite{kawashima96,alex-stiff}, but the 2d model orders only
at $T=0$ \cite{kawashima97}. 
So the question arises whether it is possible to detect this difference in
the structures of the ground-state landscape as well. Several results
using direct calculations of ground states of 2d
spin glasses using approximation methods are known 
\cite{cheng83,swendson88,freund89,berg92,sutton94,gropengiesser95}. 
Also exact ground-states 
up to $L=50$ have been analyzed \cite{kawashima97,simone96}. 
But in all these studies
only one ground state per realization was calculated, so it was  not possible
to study the structure of degeneracy. 
First attempts of the calculation of many
different ground states can be found in \cite{stauffer78,alex2}.

In the present study many different and independent true ground states
of 2d $\pm J$ spin glasses are examined in detail. 
A combination of a genetic algorithm
\cite{pal96,michal92} and  {\em Cluster-Exact Approximation}  
(CEA) \cite{alex2} is used. Recently it has been shown that this method
is able to calculate true ground states of spin glasses \cite{alex-stiff}. 
Since true ground states are calculated directly, one does not
encounter ergodicity problems or critical
slowing down like in algorithms which base on Monte-Carlo methods.
The ground-state
landscape is investigated and it is discussed whether the MF or the DS
picture describes its structure better.  Also the ground states
are analyzed whether they are ordered in an ultrametric fashion.

The paper is organized as follows:
 at first the algorithm used here for the
calculation of the ground states is explained. Next all observables
are defined. Then the results
are presented. In the last section a conclusion is driven.
 
\section*{Algorithm}

For readers not familiar with the calculation of spin-glass ground states now
a short introduction to the subject and a description of the algorithm used
here are given. A detailed overview can be found in \cite{rieger98}

\ifthenelse{\equal{\mystyle}{pp}}{
\figK
}{}

The concept of  {\em frustration} \cite{toulouse77} is important for
understanding the behavior of $\pm J$  Ising spin glasses. 
The simplest example of a frustrated system is a triple
of spins where all pairs are connected by antiferromagnetic bonds,
see fig. \ref{fig_triangle}. 
A bond is called {\em satisfied} if it contributes with a negative value
to the total energy by
choosing the values of its adjacent spins properly.
For the triangle it is
not possible to find a spin-configuration were all bonds are satisfied.
In general a system is frustrated if closed
loops of bonds exists, where the product of these bond-values is negative.
For square and cubic systems the smallest closed loops consist of four
bonds. They are called (elementary) {\em plaquettes}.

As we will see later the presence of frustration makes the 
calculation of exact ground states
of such systems computationally hard.
Only for the special case of the two-dimensional system with
periodic boundary conditions in no more than one direction and without
external field a polynomial-time algorithm is known
\cite{barahona82b}.  Now  for the general case three basic methods 
are briefly reviewed  and the
largest system sizes which can be treated are given 
for three-dimensional systems, the standard
spin-glass model.

The simplest method works by
enumerating all $2^N$ possible states and has obviously an exponential
running time. Even a system size of $4^3$ is too large.  The basic idea 
of the so
called {\em Branch-and-Bound} algorithm \cite{hartwig84} is to exclude
the parts of the state space, where no low-lying states can be found, so
that the complete low-energy landscape of systems of size $4^3$ 
can be calculated \cite{klotz}. 

A more
sophisticated method called {\em Branch-and-Cut} \cite{simone95,simone96}
works by rewriting the quadratic energy function  as a linear function
with
an additional set of inequalities which must hold for the feasible solutions.
Since not all inequalities are known a priori the method iteratively
solves the linear problem, looks for inequalities which are violated,
and adds them to the set until the solution is found. Since the number
of inequalities grows exponentially with the system size the same
holds for the computation time of the algorithm. With Branch-and-Cut
anyway small systems
up to $8^3$ are feasible.

 The method used here is able to
calculate true ground states \cite{alex-stiff} up to size $14^3$. 
For two-dimensional systems sizes up to $50^2$ can be
treated. This
is about the same size Branch-and-Cut can solve, but in contrast to that
method the algorithm used here is able to calculate many independent
ground states for each realization of the randomness.
The method bases on a special genetic
algorithm \cite{pal96,michal92} and on  Cluster-Exact Approximation  
\cite{alex2}. CEA is  an optimization method designed specially 
for spin glasses. Its basic idea is to transform the spin glass in a
way that graph-theoretical methods can be applied, which work only for systems
exhibiting no frustrations.
Next a description  of the genetic CEA is given.

Genetic algorithms are biologically motivated. An optimal
solution is found by treating many instances of the problem in
parallel, keeping only better instances and replacing bad ones by new
ones (survival of the fittest).
The genetic algorithm starts with an initial population of $M_i$
randomly initialized spin configurations (= {\em individuals}),
which are linearly arranged using an array. The last one is also neighbor of
the first one. Then $n_o \times M_i$ times two neighbors from the population
are taken (called {\em parents}) and two new configurations called
{\em offsprings} are created. For that purpose the {\em triadic crossover}
is used which turned out to be very efficient for spin glasses: 
a mask is used which is a
third randomly chosen (usually distant) member of the population with
a fraction of $0.1$ of its spins reversed. In a first step the
offsprings are created as copies of the parents. Then those spins are selected,
 where the orientations of the
first parent and the mask agree \cite{pal95}. 
The values of these spins
are swapped between the two offsprings. Then a {\em mutation}
 with a rate of $p_m$
is applied to each offspring, i.e. a fraction $p_m$ of the
spins is reversed.

Next for both offsprings the energy is reduced by applying
CEA:
The method constructs iteratively and randomly 
a non-frustrated cluster of spins.
Spins adjacent to many unsatisfied bonds are more likely to be added to the
cluster. During the construction of the cluster a local gauge-transformation
of the spin variables is applied so that all interactions between cluster
spins become ferromagnetic.

\ifthenelse{\equal{\mystyle}{pp}}{
\figA
}{}

Fig. \ref{fig_cea_example} shows an example of how the construction 
of the cluster works using a small spin-glass system.
For 2d $\pm J$ spin glasses each cluster
contains typically 70 percent of all spins.
The  non-cluster spins act like local magnetic fields on the cluster spins,
so the ground state of the cluster is not trivial.
Since the cluster has only ferromagnetic interactions, 
an energetic minimum state for its spins can be  calculated in polynomial time
by using graph theoretical methods 
\cite{claibo,knoedel,swamy}: an equivalent network is constructed
\cite{picard1}, the maximum flow is calculated 
\cite{traeff,tarjan}\footnote{Implementation details: We used 
Tarjan's wave algorithm together
with the heuristic speed-ups of Tr\"aff. In the construction of 
the {\em level graph} we allowed not only edges $(v, w)$
with level($w$) = level($v$)+1, but also all edges $(v,t)$ where $t$
is the sink. For this measure, we observed an additional speed-up of
roughly factor 2 for the systems we calculated.} and the spins of the
cluster are set to their orientations leading to a minimum in energy. 
This minimization step
is performed $n_{\min}$ times for each offspring.

Afterwards each offspring is compared with one of its parents. The
pairs are chosen in the way that the sum of the phenotypic differences
between them is minimal. The phenotypic difference is defined here as the
number of spins where the two configurations differ. Each
parent is replaced if its energy is not lower (i.e. not better) than the 
corresponding offspring.
After this whole step is done $n_o \times M_i$ times, the population
is halved: From each pair of neighbors the configuration 
 which has the higher energy is eliminated. If more than 4
individuals remain the process is continued otherwise it
is stopped and the best individual
is taken as result of the calculation.

The representation in fig. \ref{fig_algo} summarizes the algorithm. 

The whole algorithm is performed $n_R$ times and all configurations
which exhibit the lowest energy are stored, resulting in $n_G$ statistically
independent ground-state configurations.

This algorithm was already applied to examine the ground-state 
landscape of 3d spin glasses \cite{alex-sg2}.

\ifthenelse{\equal{\mystyle}{pp}}{
\figALGO
}{}

\section*{Observables}

For a fixed realization $J=\{J_{ij}\}$ of the exchange interactions and two
replicas
$\{\sigma^{\alpha}_i\}, \{\sigma^{\beta}_i\}$, the overlap \cite{parisi2}
is defined as

\begin{equation}
q^{\alpha\beta} \equiv \frac{1}{N} \sum_i \sigma^{\alpha}_i
\sigma^{\beta}_i 
\label{def_q}
\end{equation}

The ground state of a given realization is characterized by the probability
density $P_J(q)$. Averaging over the realizations $J$, denoted
by $[\,\cdot\,]_{av}$, results in ($Z$ = number of realizations)

\begin{equation}
P(q) \equiv [P_J(q)]_{av} = \frac{1}{Z} \sum_{J} P_J(q) \label{def_P_q}
\end{equation}

The probability densities are symmetric as no external field is applied:
$P_J(q) = P_J(-q)$ and $P(q) = P(-q)$. Hence, only averages of $|q|^n$ are
relevant:

\begin{eqnarray}
\overline{|q_J|^n} & \equiv & \int_{-1}^1 |q|^n P_J(q) dq \label{def_mean_q}\\
\overline{|q|^n} & \equiv & \int_{-1}^1 |q|^n P(q) dq 
\end{eqnarray}

The Droplet model predicts that only two pure states exist, implying
that $P(q)$ converges for $L \to \infty$ to 
$P(q) = \frac{1}{2}(\delta(q-q_{EA})+\delta(q+q_{EA}))$,
while in the MF picture
the density remains nonzero for a range $-q_{EA} \le q \le q_{EA}$ with
peaks at $\pm q_{\max}$ ($0< q_{\max} \le q_{EA}$, $q_{\max}\to q_{EA}$ 
for $L\to\infty$). Consequently the variance

\begin{equation}
\sigma^2(|q|) \equiv \int_{-1}^1 (\overline{|q|} - |q|)^2P(q)\,dq = 
\overline{|q|^2} - \overline{|q|}^2 \label{def_sigma_q}
\end{equation}

stays finite for $L \to \infty$ in the MF pictures 
($\sigma^2(|q|) = s_{\infty}+s_1/L^{s_2}$) \cite{berg94}, while 
$\sigma^2(|q|) \sim L^{-y} \to 0$ according the DS approach. Here, $y$ is the
zero-temperature scaling exponent \cite{mcmillan}, which is denoted as
$\Theta$ in  \cite{fisher1,fisher2}.

Another way of describing the finite-size behavior of $P(|q|)$ is to
sum up the contributions from small overlap-values $q\le q_0$:

\begin{equation}
X_{q_0} \equiv \int_{-q_0}^{q_0}P(q)\,dq \label{def_xq0}
\end{equation}
This value should converge to 0 in the DS picture as long as $q_0<q_{EA}$
while it should stay non-zero for the MF framework.

The overlap defined in (\ref{def_q})
can be used to measure the distance $d^{\alpha\beta}$ between two states:

\begin{equation}
d^{\alpha\beta} \equiv 0.5(1-q^{\alpha\beta})
\end{equation}
with $0\le d^{\alpha\beta} \le 1$. 
For three replicas $\alpha,\beta,\gamma$ the usual triangular inequality
reads
$d^{\alpha\beta} \le d^{\alpha\gamma} + d^{\gamma\beta}$.
Written in terms of $q$ it becomes

\begin{equation}
q^{\alpha\beta} \ge q^{\alpha\gamma} + q^{\gamma\beta}-1
\label{triangular_q}
\end{equation}

In an {\em ultrametric} space \cite{rammal86} 
the triangular inequality is replaced by a stronger one
$d^{\alpha\beta} \le \max(d^{\alpha\gamma}, d^{\gamma\beta})$
or equivalently

\begin{equation}
q^{\alpha\beta} \ge \min(q^{\alpha\gamma}, q^{\gamma\beta})
\label{ultra_q}
\end{equation}

An example of an ultrametric space is the set of leaves of a binary tree:
the distance between two leaves is defined by the number of edges on a path
between the leaves.

Let $q_1\le q_2 \le q_3$ be the overlaps $q^{\alpha\beta}$, 
$q^{\alpha\gamma}$, 
$q^{\gamma\beta}$ ordered according their sizes.
By writing the smallest overlap on the left side in equation (\ref{ultra_q}), 
one realizes that two of the overlaps must be equal and 
the third may be larger or the same: $q_1 = q_2 \le q_3$

In a finite-size system this relation may be violated.
Here two ways are used of determining whether ground states of
realistic spin glasses
become more and more ultrametric with increasing size $L$:

\begin{itemize}
\item
The difference 
\begin{equation}
\delta q\equiv q_2-q_1
\label{def_delta}
\end{equation}
 is calculated for all triplets. 
Because the influence of the absolute size
of the overlaps should be excluded the third overlap is fixed: 
$q_3=q_{fix}$. In practice 
only overlap triples are used where $q_3 \in [q_{fix},q_{fix2}]$ holds
to obtain sufficient statistics . With
increasing size $L$ the distribution $P(\delta q)$ should tend  to a 
Dirac delta function for
an ultrametric system \cite{bhatt86}.
\item
If two overlaps are fixed ($q^{\alpha\gamma}=q^{\beta\gamma}=q_{fix}$,
in practice $q^{\alpha\gamma},q^{\beta\gamma}\in [q_{fix},q_{fix2}]$), 
equation (\ref{triangular_q}) implies $q\equiv q^{\alpha\beta}\ge 2q_{fix}-1$
while ultrametricity implies $q\ge q_{fix}$ which is stronger if $q_{fix}<1$
\cite{cacciuto97}. The
distribution $P_{2-fix}(q)$ of the third overlap is used to characterize
the ultrametricity of a system. 
Additionally the weighted 
fraction of the distribution outside $[q_{fix},q_{EA}]$ 
\begin{eqnarray}
I_L & \equiv & \int_{-1}^{q_{fix}}P_{2-fix}(q)(q-q_{fix})^2\,dq \\\nonumber
& & + \int_{q_{EA}}^{1}P_{2-fix}(q)(q-q_{EA})^2\,dq
\label{def_I_L}
\end{eqnarray}
(see \cite{cacciuto97}) 
should vanish for $L\to\infty$ in an ultrametric system.
\end{itemize}

\section*{Results}
We used  simulation parameters determined in the following way:
For each system size  several different combinations
of the 
parameters $M_i, n_o, n_{min}, p_m$  were tested. 
For the final parameter sets it is not possible to obtain lower
energies even by using parameters where the calculation consumes  
four times the computational effort.
Using parameter sets chosen this way genetic CEA calculates true
ground states, as shown in  \cite{alex-stiff}. 
Here a mutation rate of $p_m=0.05$ and $n_R=40$ runs per
realization were used for all system sizes.
Table 1 summarizes the parameters and gives the
typical computer time $\tau$ spent per ground state
computation on a 80 MHz PPC601. 
For each system size ground states for 1000 different realizations of the
disorder were calculated.
On average $n_G > 29$ (not necessarily different)
ground-state configurations 
were obtained for all system sizes $L$ using  $n_R=40$ runs per realization.

\ifthenelse{\equal{\mystyle}{pp}}{
\tabA
}{}

\ifthenelse{\equal{\mystyle}{pp}}{
\figB
}{}

At first the result for the ground-state energy as function of system
size is presented in fig. \ref{fig_e_min}. By performing a
finite-size scaling (FSS) analysis using the function $e(L)=e_{\infty}+
e_1L^{-e_2}$ a ground-state energy of the infinite system of
$e_{\infty}=-1.4015(3)$ is obtained. This is consistent with an extrapolation
from exact ground states of finite systems $e_{\infty}=-1.4015(8)$ 
\cite{simone96} and with other former results from transfer-matrix calculations
$e_{\infty}=-1.402(1)$ \cite{cheng83}, Monte-Carlo simulations
 $e_{\infty}=-1.407(8)$ \cite{swendson88}, multicanonical simulations
 $e_{\infty}=-1.394(7)$ \cite{berg92}, genetic algorithms
 $e_{\infty}=-1.400(5)$ \cite{sutton94},  $e_{\infty}=-1.401(1)$ 
\cite{gropengiesser95}, a
special cluster-construction method $e_{\infty}=-1.402(2)$ \cite{freund89}  
and pure CEA  $e_{\infty}=-1.400(5)$ \cite{alex2}. The value presented
here has a higher accuracy than the former results.

The exponent of the decay is $e_2=2.2(1)$ which is much larger than the
upper limit of $y\equiv e_2\le (d-1)/2=0.5$ proposed by DS. 
This is an evidence that
the DS picture may not be appropriate for 2d realistic spin glasses.

Information about the ground-state landscape 
can be extracted by evaluating the distribution
of overlaps. In fig. \ref{fig_PLq}  $P(|q|)$ 
is shown for three sample sizes $L=10,20,40$. The distribution
is averaged over the disorder, where each realization enters the
result with the same weight, independent of the number of ground-state
configurations which were available. The distributions extent over large
intervals down to $q=0$ which indicates the existence of a complex
ground-state landscape. For small overlaps no large change is visible 
with increasing system size, but the peak of the distributions located at
larger q-values shifts to smaller values.

\ifthenelse{\equal{\mystyle}{pp}}{
\figC
}{}

\ifthenelse{\equal{\mystyle}{pp}}{
\figI
}{}

The position $q_{\max}$ of this peak can be used to calculate the
Edwards-Anderson order parameter $q_{EA}$ which is the maximum value
of $q$ where $P(q)$ is nonzero 
in the infinite system. Fig. \ref{fig_q_max} shows
the value of $q_{\max}$ as function of $L$. Using a FSS
fit with $q_{\max}(L)=q_{EA}+q_1 L^{-q_2}$ a value of $q_{EA}=0.50(9)$ is
obtained. The resulting function is shown in the figure using a line.

\ifthenelse{\equal{\mystyle}{pp}}{
\figD
}{}

So far we have seen that finite systems exhibit a complex
ground-state landscape.
But to decide whether this is true even for the infinite system
one must investigate the shape of $P(|q|)$
as function of system size $L$. In fig. \ref{fig_sigma} the variance
(see def. (\ref{def_sigma_q})) of the distribution is shown as function
of $L$. By fitting the variance to a function of form 
$\sigma ^2(L)=s_{\infty} +s_1L^{-s_2}$
a value of $s_{\infty}=0.004(8)$ was obtained. This result is very close
to zero. In the figure a fit with 
$s_{\infty}\equiv 0$ is shown which looks very reasonable. 
So it is possible that for $L\to\infty$
the width of the distribution shrinks to zero, which would mean that
the ground-state landscape is simple, as described by the DS framework.
This is different from the case of the three-dimensional model, 
where the same procedure
resulted in $s^{3d}_{\infty}=0.0608(6)$ \cite{alex-sg2}.

However, the impression mediating from fig. \ref{fig_PLq} is different: a long
tail down to $q=0$ persists for all system sizes, so a width of zero
for the infinite systems seems unlikely. 
Consider a infinite system, where the overlaps are distributed
according a distribution
with a constant probability-density of $0.5$ for $q\le 0.5$ and a 
delta-function at $q=0.5$ with weight $0.75$: $P(q)=0.25 \Theta(0.5-p)
+0.75 \delta(p-0.5) \,\,(p\ge 0)$, which seems plausible from fig.
\ref{fig_PLq} and \ref{fig_q_max}. Then one obtains a variance 
$\sigma^2(|q|)=0.017$
which is very close to $s_{\infty}$ regarding the given error-bar. 

Since the result is not definite so far, next the contribution of 
small overlap-values to $P(|q|)$ is studied.
The fraction $X_{q_0}$ (see def. (\ref{def_xq0})) of the distribution 
below a given value $q_0$ is displayed
in fig. \ref{fig_x02} using $q_0\equiv 0.2<0.5=q_{EA}$ as function
of system size. As $X_{q_0}$ seems to be independent of the system
size it is reasonable to conclude that the infinite system has a broad
distribution.

So far the question whether the ground-state landscape is complex
has been addressed: a complex landscape for the 2d $\pm J$ 
spin glass seems likely, but the
results are less definite than earlier results for the 3d model. 

\ifthenelse{\equal{\mystyle}{pp}}{
\figE
}{}

\ifthenelse{\equal{\mystyle}{pp}}{
\figF
}{}

In the second part of this section it is studied
whether the ground states are ultrametrically ordered. For that purpose
for each realization all possible triplets of ground states were
chosen and the corresponding three overlaps evaluated. The
the quantity $\delta q$ which is the difference between the two
smaller
overlap values was calculated (see equation (\ref{def_delta})) for
all possible triplets with a constraint for the largest of the three
overlaps: $q_3 \in [\frac{3}{5}q_{EA}-0.05,\frac{3}{5}q_{EA}+0.05]$ =
$[0.25,0.35]$ (which implies $\delta q < 0.35$ as well). 
For an ultrametric system the distribution of $\delta
q$ should be a delta function.
To improve the statistics
we used the absolute value of all overlaps. The distribution $P(\delta q)$
is shown in fig. \ref{fig_deltaq} for $L=10,20,40$. Each realization enters
the distribution with the same weight. One can see that the 
distribution is independent of the system size. The
average value of
$\delta q$ as function of system size shown in the inset. Hence,
in the infinite system $\delta q>0$ is possible.   
It seems that the 2d $\pm J$ spin glass is not ultrametric, which
is in strong contrast to the result for the three-dimensional model
where $P(\delta q$) converges to a delta-function \cite{alex-ultra}.

\ifthenelse{\equal{\mystyle}{pp}}{
\figG
}{}

Additional information can be obtained
by fixing two of the three overlaps of a triplet. 
We took all triplets where two arbitrary 
overlaps fell into
the interval $[\frac{3}{5}q_{EA}-0.05,\frac{3}{5}q_{EA}+0.05]$. The 
resulting distributions $P_{2-fix}(q)$  of the
third remaining overlap is shown in fig. \ref{fig_P_q_2fix} for $L=10,20,40$.
The triangular inequality gives $q>-0.5$ whereas for an ultrametric
system $q>0.25$ must hold. We concentrate on the part of the
distribution with $q<0.25$. Although the shape of the 
distribution changes a little bit,
the fraction of the overlaps forbidden in an
ultrametric system keeps fairly constant:
the inset shows the fraction $I_{2-fix}(0.25)$ of the
distribution below $q=0.25$.  Again reasonable evidence 
against an ultrametric organization of the ground-states is found, but
it is a little bit weaker, since very small values near $q=-0.5$ disappear
for large systems.

\ifthenelse{\equal{\mystyle}{pp}}{
\figH
}{}

To complete the comparison with \cite{alex-ultra}, the distribution
$P_{2-fix}(q)$
integrated outside $[0.25,0.5]$ (see def. of $I_L$ in (\ref{def_I_L}))
is shown in fig. \ref{fig_I_L}. Again the result is different from the
case of the three-dimensional $\pm J$ spin glass: 
here $I_L$ seems to remain non-zero for $L\to\infty$. 

\section*{Conclusion}

Many different and independent ground states for 2d $\pm J$ spin glasses
were calculated up to sizes $L=40$ using the genetic 
Cluster-Exact Approximation. 
From former calculations and comparison with exact results is it 
clear that true ground states were obtained.

By evaluating the distribution of overlaps
evidence for a complex organization of the ground
states  is found, but the evidence is weak, since it is not
clear whether the width of the distribution of overlaps scales to
zero with increasing system size. This is similar to the 3d case, but there
the evidence for the MF picture is stronger.

By studying triplets of ground states with one or two of the
three overlap values fixed
convincing evidences are found that the ground states of 2d $\pm J$ 
spin glasses are
organized not in an ultrametric manner. This is very different from the
three-dimensional model. To our opinion this may be connected with
the fact that 3d
realistic spin glasses have an ordered phase at finite temperature
whereas the 2d model shows a spin-glass phase only at $T=0$.

Since only moderate system sizes up to $N=40^2$  were investigated,  the
behavior might change at larger sizes. But the finite-size behavior
of the data presented here is very smooth, thus, we
believe that our results persist for $L\to\infty$.

The results presented here were obtained for a bimodal distribution
of the interactions.
For spin glasses with Gaussian distribution
the ground state is unique, implying that $P_L(|q)|)$ are both the
same in the MF
and DS picture. The predictions are different at finite
 temperature. We expect the same behavior as for the $\pm J$ model
if one allows deviations of order one
from the true ground-state energy, but for that system type ground states are
much harder to calculate using the genetic CEA algorithm.

\section*{Acknowledgements}

The author thanks H. Horner and G. Reinelt for manifold support.
He thanks K. Bhattacharya for critical reading of the manuscript 
and for giving  many helpful hints.
The author took much benefit from discussions with S. Kobe, H. Rieger
and A.P. Young.
He is also grateful to the {\em Paderborn Center for Parallel Computing}
 for the allocation of computer time. This work was supported
by the Graduiertenkolleg ``Modellierung und Wissenschaftliches Rechnen in 
Mathematik und Naturwissenschaften'' at the
{\em In\-ter\-diszi\-pli\-n\"a\-res Zentrum f\"ur Wissenschaftliches Rechnen}
 in Heidelberg.

\ifthenelse{\equal{\mystyle}{sub}}{

\clearpage
{\bf Captions}

\begin{enumerate}
\item \captionK
\item \captionA
\item \captionALGO
\item \captionB
\item \captionC
\item \captionI
\item \captionD
\item \captionE
\item \captionF
\item \captionG
\item \captionH
\end{enumerate}

\clearpage
\onecolumn
{\bf Figures}\\[1cm]
\figK

\clearpage
\onecolumn
\figA

\clearpage
\onecolumn
\figALGO

\clearpage
\onecolumn
\figB

\clearpage
\onecolumn
\figC

\clearpage
\onecolumn
\figI

\clearpage
\onecolumn
\figD

\clearpage
\onecolumn
\figE

\clearpage
\onecolumn
\figF

\clearpage
\onecolumn
\figG

\clearpage
\onecolumn
\figH

\clearpage
{\bf Table}\\[1cm]
\tabA

}{}

\end{document}